# A BRIEF SUMMARY OF ELECTROMAGNETIC QUANTUM GRAVITY

Tom Ostoma and Mike Trushyk
Email: emgg@rogerswave.ca
Monday, February 22, 1999

ABSTRACT:

We briefly review the current status of a new quantum gravity theory called Electro-Magnetic Quantum Gravity (ref. 1) or EMQG introduced several months ago. EMQG is manifestly compatible with Cellular Automata (CA) theory (ref. 2 & 4), and is based on a new theory of inertia (ref. 5) proposed by R. Haisch, A. Rueda, and H. Puthoff. Newtonian Inertia is due to the strictly local, electrical force interactions of matter particles (consisting of real electrically charged fermions) with the surrounding, electrically charged, virtual fermion particles of the quantum vacuum. The force originates from each charged fermion particle of the mass undergoing relative acceleration with the the quantum vacuum particles. The sum of all these tiny electrical forces originating from each electrically charged particle in the mass is the source of the <u>total inertial force</u> of a mass, which is the force that opposes accelerated motion in Newton's famous inertia law 'F = MA'. Gravity also involves the <u>same 'inertial' electrical force component</u> that exists for inertial mass described above. The Weak Equivalence Principle turns out to be a physical phenomenon, originating from common 'lower level' quantum processes in both gravitational and inertial mass. The magnitude of the gravitational mass of a test mass on the earth results from the same quantity of electrical force interactions as in inertia, but on the earth it is the virtual fermions of the quantum vacuum that are <u>actually accelerating</u> downward (on a statistical average basis) with respect to the test mass.

In large gravitational fields like the earth, 4D curved space-time is a consequence of the behavior of light and matter under the influence of the (statistical average) downward accelerated 'flow' of <u>electrically charged</u> virtual fermion particles of the quantum vacuum. Photons from a light source scatter through the influence of the downward accelerating, virtual <u>charged</u> particles of the vacuum, thus deflecting the photons from their normal straight line paths. However at minute quantum distance scales 4D space-time does not exist, and one only sees quantum vacuum virtual particles interacting with real matter particles which all live in a quantized (flat) CA space and separate quantized time. The coordinated 'accelerated flow' of the quantum vacuum virtual particles can be visualized as a special 'Fizeau fluid' that flows through everything that is subjected to a gravitational field. This same process also occurs to masses and light subjected to accelerated motion. Like the Fizeau experiment (which was performed with flowing water) the behavior of photons, clocks, and rulers are now affected by the <u>accelerated</u> flow of the charged virtual particles of the quantum vacuum.

## 1. BRIEF SUMMARY OF EMQG THEORY

We present a brief review of a new quantum gravity theory called Electro-Magnetic Quantum Gravity or EMQG (ref. 1). EMQG provides several new paradigms for physical reality. First, EMQG is based on the idea that our universe is a vast numeric Cellular Automata (CA) simulation, which is the most massively parallel computer model known to science (references 2 and 4). This CA structure proposed in EMQG is a simple 3D geometric CA. All physical phenomena, including space, time, matter, and forces are the result of the interactions of *information* patterns, or particles governed by the mathematical laws and the connectivity of the CA. Because of the way the CA functions, all the known global laws of the physics must result from the activities of the local mathematical law that governs each cell, where each and every cell contains the *same* mathematical law or program.

The CA structure places severe constraints on the nature of physical reality. The CA structure automatically dictates that our universe *must* have a *maximum speed limit* for any type of motion (the transfer of information patterns from cell to cell, ref. 3). This limitation results from the way that a cell is allowed to change state. It can only change state on the next clock interval based on the numeric contents of it's immediate neighbors and the rules programmed in that cell. Information must transfer sequentially from cell to cell at some maximum rate determined by the quanitization scale of the CA. For example, if the plank length $L_p$ represents the smallest distance increment (or the shifting of information in any direction by one cell) and the plank time $T_p$ represents the smallest time interval (1 'clock cycle'), then the fastest that information can be moved from place to place by our measure is $L_p / T_p$ , which is equal to the velocity of light 'c' or the plank velocity.

All phenomena in the universe must result from the activities of the information patterns, which roam around in the CA quantized space. This must include all the known forces. Quantum field theory requires that all forces result from a boson particle exchange process. We found that the particle exchange paradigm fits naturally within the CA theory structure, where the boson exchange represents the transfer of boson information patterns between (fermion) matter particles, resulting in changes in the motion of the receiving particle. In EMQG, we have maintained that *all* forces (gravity is no exception) originate from a exchange processes, as dictated by quantum field theory. However, we found that the effects of gravity result from *two* pure, boson force exchange processes. **Both** the photon and graviton exchanges occurring simultaneously inside a large gravitational field like the earth. Photon exchanges between the quantum vacuum and the charged particles of the mass is *primarily responsible* for the magnitude of the mass. The graviton exchanges between the earth and the quantum vacuum cause the virtual particles of the quantum vacuum to fall during their short lifetimes. Graviton exchanges do occur directly between the earth and the test mass, but the resulting force is negligible by 40 orders of magnitude, in comparison to the quantum vacuum electrical forces.

We modified a new theory of inertia first introduced by R. Haisch, A. Rueda, and H. Puthoff in 1994 (ref. 5), which we refer to as HRP inertia. In HRP inertia, these authors claim that inertia, or the opposition force of a mass to any acceleration, is caused by the resulting electrical forces between particles in a test mass and the surrounding virtual particles of the quantum vacuum. In particular, they claim that it is the Lorentz force from each of the charged 'parton' particles making up a mass with the background virtual photon field (which they called the zero point fluctuations or ZPF). The parton is an old fashion term for the constituent particles that make up the nuclear particles, such as protons and neutrons, first coined by R. Feynman.

We were forced to modified HRP inertia based on our study of the principle of equivalence, and the modified theory is called Quantum Inertia (or QI), which formas an integral part of EMQG theory. This modification of HRP inertia involves the introduction of new particle of nature which we call the masseon particle. Masseons readily combine together to form all the known (fermion) mass particles. The masseon is electrically charged, as well as possessing mass-charge. We define mass-charge as the ability of a particle to emit and absorb graviton particles, the carrier of the pure gravitational force.

Quantum Inertia is based on the idea that inertial force is due to the tiny electrical force interactions (where the exact details are not fully developed at this time) originating from each charged masseon particle that makes up real matter undergoing relative acceleration with respect to the virtual, electrically charged masseon particles that consist of the quantum vacuum. These tiny forces are the origin of the total resistance force to accelerated motion in Newton's law 'F = MA', where the **sum of each of the tiny masseon electrical forces totals to the total inertial force in Newton's law**. The exact details of the tiny electrical force process is not fully known, and hence this remains a postulate of EMQG. The main difference between HRP inertia and Quantum Inertia is that the quantum vacuum background that is involved in this process is due to virtual fermions, rather then the virtual photons (or ZPF) as in the original paper. This change was necessary in order to understand the principle of equivalence at the quantum level.

This new quantum approach to classical inertia automatically resolves the problems and paradoxes of accelerated motion introduced in Mach's principle, by suggesting that the virtual particles of the quantum vacuum serve as Newton's universal reference frame (which Newton called absolute space) for accelerated motion only. Thus, Newton was correct in his statement that it is the relative accelerated motion with respect to absolute space that somehow determines the quantity of inertia of a mass. This statement is now be reworded to state that it is relative acceleration of (the particles that make up) a mass with respect to the (net statistical average) accelerated state of the virtual masseons of the quantum vacuum that causes the force of inertia. However, unlike the $19^{th}$ century classical ether, the *quantum vacuum reference frame* **cannot** be used determine your absolute velocity in space, and hence the quantum vacuum cannot be considered as a new $21^{st}$ century 'ether'. In fact, it is well known experimentally that it is impossible to determine the *absolute* velocity of a test mass. This statement is one (of the two) founding postulates of special relativity.

The relative accelerated motion of the virtual masseons of the quantum vacuum with respect to the average motion of the real masseon particles that are bound in a mass can also be used as the reference frame to define absolute mass (which is equivalent to the special relativistic rest mass). This is contrary to the well known mass-velocity formula of special relativity (ref. 3). However, in EMQG the mass-velocity formula *is still valid*, but now results from the *reduction in the effectiveness of the force* between two very high velocity reference frames.

To see this, one must recall the particle exchange paradigm, where the quantity of force transmitted between two objects in different inertial reference frames depends on the flux rate of the exchange particles. In other words, the number of particles exchanged per unit of **time** represents the magnitude of the force transmitted between the two particles. If the relative velocity is low or v<<c, then the exchange process appears almost the same as when the two particles are at relative rest. This is because the velocity of light, which is the speed of the exchange particles, is very high compared to the relative velocity of the two particles, and the flux rate remains largely unaffected by the motion.

Now as the relative recession velocity of the two frames approaches light velocity, it is comparable to the velocity of the exchange particle, and this effects the received flux rate. This can be seen from the Lorentz time dilation, and the arrival times of the exchange particle is altered as given by $t = t_0 / (1 - v^2/c^2)^{1/2}$. The timing of the exchange particles is altered, and therefore the flux rate is altered as well, since flux has units of numbers of exchange particles per unit time. Therefore, the magnitude of the force is reduced by: $F = F_0 (1 - v^2/c^2)^{1/2}$; where $F_0$ is the magnitude of the force when at relative rest. Thus, we have concluded that the force is actually reduced in strength, and *not* that the mass has increased. The reduced force (say the electrical force used in a linear particle accelerator for example) is less effective in accelerating a relativistic charged mass. This phenomena is then mistaken as an actual effective increase of the mass of the accelerated charged object (an electron in a particle accelerator for example).

We found that gravity also involves the same 'inertial' electrical force component that exists for an accelerated mass, which reveals a *deep connection* between inertia and gravity. Inside a large gravitational field there exists a similar quantum vacuum process that occurs for inertia, where the roles of the real charged masseon particles of the mass and the virtual electrically charged masseons of the quantum vacuum are reversed. Now it is the electrically charged virtual particles of the quantum vacuum that are accelerating, and the mass particles that are at relative rest. Furthermore, the general relativistic Weak Equivalence Principle (WEP) results from this common physical process existing at the quantum level in both gravitational mass and inertial mass. Figures 2A - 2D illustrate this action.

What is unique about EMQG theory is that gravity involves **two force exchange particles**! In gravity, *both* the electrical force (photon exchange process) and the pure gravitational force (graviton exchange process) interactions are occurring simultaneously for all gravitational interactions. Graviton particles originate from all the mass particles (masseons) in the earth in vast numbers, and are absorbed by any particle possessing mass. If one believes that the some of the virtual particles of the quantum vacuum consists of particles that possess mass (virtual fermions, or virtual masseons in EMQG), then one must conclude that these virtual particles are in a state of free fall (downward acceleration) during their very short lifetimes near the earth. We have seen that the electrical interaction of real matter particles with the quantum vacuum particles is the source of the electrical force interaction in inertia when a state of relative acceleration exists. Therefore, the falling vacuum causes a stationary test mass on the earth to exert a force against the earth's surface, which we call the weight of that mass. Furthermore, we can now easily see why the weight of a mass on the earth is equal to the inertial mass of the same object on the floor of a rocket accelerating at 1g. Figures 2A-2D illustrate the interaction between the vacuum and matter where a state of relative acceleration exists between the two.

However for a gravitational test mass, the graviton exchange process (which is totally negligible in inertial reference frames) occurring directly between the earth and the test mass upsets the perfect equivalence of inertial and gravitational mass, where the gravitational mass is slightly larger than the inertial mass due to the extra force from the direct graviton exchanges. One of the consequences of this imbalance is that if a very large and a tiny mass are dropped simultaneously on the earth, the larger mass would arrive slightly sooner. Since this is in violation of the WEP, the strong equivalence principle is no longer applicable. This effect might be measurable in the laboratory.

In EMQG, we proposed that all (fermions) matter particles get their quantum mass numbers from combinations of just <u>one</u> fundamental matter (and anti-matter) particle called the 'masseon' particle. The masseon has one fixed, (lowest) quanta of mass, which we called low level 'mass charge'. The masseon also carries either a positive or negative (lowest) quanta of electric charge. Furthermore, we proposed a new universal constant "i", defined as the inertial force existing between the quantum vacuum and a single charged masseon particle, which is accelerating at 1 g. This force represents the lowest possible quanta of inertial force (at 1g) and gravitational force on the earth. The masseon particle generates a fixed flux of gravitons (in analogy to electrical charge), with the flux rate being <u>unaffected</u> by relativistic motion. In EMQG, graviton exchanges are physically similar to photon exchanges in QED, with the same concept of positive and negative gravitational 'mass charge' carried by masseons and anti-masseons, with the ratio of the graviton to photon exchange force being $10^{-40}$ . We found that graviton exchanges occur between a large mass and the surrounding virtual particles of the quantum vacuum, and they also directly occur between the large mass and a test mass. The electrical force (<u>photon exchanges</u>) between the virtual particles and the test mass (occurring in inertial frames and in gravitational frames) is responsible for the equivalence of inertial and gravitational mass. The pure gravitational force (<u>graviton exchanges</u>) is responsible for the distortion of the (net statistical average) acceleration vectors of the virtual particles of the quantum vacuum near the earth (with respect to the earth).

We also found that because there are equal numbers of virtual masseon and anti-masseon particles existing in the quantum vacuum everywhere (at any given instant of time) the cosmological constant must be very close to zero. This is the solution to the famous cosmological constant problem.

Since the state of the electrically charged virtual masseons of the quantum vacuum are very important in considerations of inertia and gravitation (and is responsible for the equivalence principle), we found that there is a deep connection between the quantum vacuum and the fact that space-time is curved in accelerated frames and gravitational frames. We introduced a new paradigm for the origin of 4D, curved Minkowski space-time near a large mass. We found that 4D space-time is simply a consequence of the behavior of matter (fermions) and energy (photons) under the influence of the (net statistical average) downward accelerated 'flow' of the charged virtual particles of the quantum vacuum.

The state of accelerated 'flow' of the virtual particles of the quantum vacuum can be thought of as a special 'Fizeau-like fluid' (unknown to Einstein when he was developing relativity). Like in the Fizeau experiment (performed with the flow of constant velocity water) the behavior of light, clocks, and rulers are now affected by the <u>accelerated</u> 'flow' of the virtual particles of the quantum vacuum with respect to a large mass (and in accelerated frames). This accelerated flow can now **act** on the motion of matter and light, to distort space and time. Furthermore, we have shown mathematically that the amount of space and time curvature based on EMQG corresponds to the same amount predicted by the Schwartzchild metric for the earth. This conclusion was based on the concept that photons scatter off the virtual particles of the quantum vacuum, thus maintaining the *same net acceleration* as the downward 'flow' of virtual particles (in absolute CA units). Photons, however, still move at an absolute constant speed between the virtual particle scattering.

We have concluded that the speed of light in a universe where the quantum vacuum is absent of all virtual particles is greater than the light velocity in our actual universe. This is because in our universe the real photons scatter off the virtual particles, which introduce a **small random delay** before another real photon is re-emitted. In addition, the photon undergoes second-order (and higher) scattering processes (according to QED theory), which also contribute to extra delays. Thus the velocity of light that we observe in our universe is a net average statistical value based on the 'raw low-level' photon velocity **minus** an unknown but very large velocity penalty due to the total scattering process. Over classical distance scales, the average light velocity is constant in all directions because of the immensely large number of interactions that occur, and the remarkable regularity of the quantum vacuum.

On quantum distance scales Minkowski 4D space-time gives way to the secondary (quantized) absolute 3D space and separate absolute (quantized) time required by CA theory. Curved 4D space-time is replaced by a new paradigm where curvature is a result of pure particle interaction processes. Particles occupy definite locations on the CA cells, and particle states are evolved by a universal 'clock'. All interactions are *absolute,* because they depend on absolute space and time units on the CA. However, we cannot probe this scale, because we are unable to access the absolute cell locations, and numeric contents of the cells. In this realm, the photon particle (as well as the graviton) is an information pattern, that moves (shifts) with an absolute constant 'velocity', since it merely shifts from cell to neighboring cell in every 'clock' cycle of the CA.

The relationship that exists between EMQG theory and the rest of physics is illustrated in Figure 1. Note the importance of the quantum field theory and the quantum vacuum to EMQG. Note that the principle of equivalence is derived from EMQG, and is not a postulate. In fact, the equivalence principle is not perfect!

EMQG theory has been applied to resolving some outstanding limitations and problems with Einstein's general relativity theory. These issues are addressed in the next section.

## 2. LIMITATIONS OF GENERAL RELATIVITY AND EMQG

EMQG was developed in part to overcome a number of limitations and problems with conventional general relativity theory. Most physicists assume that Einstein's theory of general relativity has no inconsistencies or flaws, and can be applied over any distance scale, or to any size and density of mass distribution. Part of this belief comes from the mathematical elegance of the theory, and partly from the spectacular experimental agreement with theory, especially over the past 25 years.

In fact general relativity is known to have a number of mathematical problems, and fails to answer some very important physical questions regarding gravitation. Furthermore, it appears that quantum theory and general relativity as currently formulated are, for the most part, completely incompatible. In spite of over 70 years of intensive research, no full theory of quantum gravity has been formulated successfully. We will briefly summarize some of these problem areas in general relativity, and show how EMQG is able to resolve these issues.

### (A) GENERAL RELATIVITY FAILS TO PROVIDE A PHYSICAL CAUSE-EFFECT OR ACTION THAT COUPLES THE QUANTITY OF MATTER-ENERGY TO THE AMOUNT OF LOCAL 4D SPACE-TIME CURVATURE.

**GENERAL RELATIVITY**: Einstein's gravitational field equations *do* dictate how much curvature of 4D space-time there is, based on how much matter-energy there is nearby. However, it does not tell you *why* there is curvature. What is the cause of 4D space-time curvature around a large mass? What is it about the quantity and density of a mass distribution that causes the local 4D space-time to curve? In general relativity there is no physical reason or action to relate matter to space-time, other then from basic considerations of the equivalence principle. In other words, curvature *just happens* in general relativity.

In fact, Einstein appealed to Newton theory of gravity to adjust the amount of coupling between matter-energy and local 4D space-time, without providing any further insights into this question. In some sense, this is like using a mathematical 'fudge' factor. The action in general relativity is adjusted so that in weak field case, the amount of curvature coincides with the expected behavior of matter in accordance to Newton's and Poisson's laws of gravity.

**EMQG:** There is definitely a physical action that couples the quantity of matter to the amount of space-time curvature. This action is caused by the graviton particle. The total amount of absolute matter determines the total number of masseon particles in the mass. Double the absolute mass, and you double the number of masseons. Double the number of masseons, and you double the flux rate of the gravitons exchanged with the virtual masseon particles in the surrounding quantum vacuum. The absorption of gravitons accelerates the virtual particles of the quantum vacuum downwards. The graviton flux spreads out from the earth, and the number of graviton particles per unit area falls as $1/4\pi r^2$, where 'r' is the distance from the center. The decrease in the graviton flux with height reduces the downward acceleration of the virtual masseon particles of the quantum vacuum with increasing height. Since the virtual masseon particles accelerating downwards also possess electrical charge, there is an electrical interaction with light and matter moving in the vicinity of the earth, which causes matter and light to move differently then in far space. We call this interaction the 'Fizeau-like' scattering from the virtual particles of the quantum vacuum, which acts like some sort of accelerated, flowing fluid medium (in analogy with the Fizeau experiment for light propagation with moving water). Therefore, the variation of the magnitude of the acceleration vectors of the virtual particles with height also represents the variation of 4D space-time curvature with the height. The action of the graviton particle is not instantaneous, however, since the gravitons move at light velocity. The finite speed of the graviton particle is largely responsible for the Lense-Thirring effect, where inertial frames seem to be dragged by the rotation of the earth. The finite speed of the graviton particle is also partially responsible for the existence of gravitational waves.

(B) **IN GENERAL RELATIVITY, THE PRINCIPLE OF EQUIVALENCE MUST BE ACCEPTED ONLY AS A POSTULATE, AND CANNOT BE DERIVED.**

**GENERAL RELATIVITY**: Why should there be equivalence between inertial and gravitational mass? After all, both these mass types are defined differently in conventional Newtonian physics. Inertial mass involves accelerated motion, but there appears to be no motion involved in the definition of gravitational mass. In general relativity, equivalence has still remained a postulate as first proposed by Einstein, in spite of 70 years of investigation, albeit a very well tested postulate.

**EMQG:** In fact, the principle of equivalence can be derived from lower level quantum particle processes, and has been investigated in detail in EMQG theory. The Equivalence

Principle and quantum inertia forms the central core of EMQG theory. We have found that equivalence arises from the *same electrical interaction* of the electrically charged virtual masseon particles of the quantum vacuum with the real electrically charged masseon particles of the test mass, which also occurs in inertia. We have also found that equivalence is *not* perfect! It breaks down when both an extremely large mass and a tiny mass are dropped on the earth, with the larger mass arriving slightly earlier.

### (C) GENERAL RELATIVITY FAILS TO ACCOUNT FOR THE ORIGIN OF INERTIA, AND CANNOT FULLY ACCOUNT FOR MACH'S PRINCIPLE.

**GENERAL RELATIVITY**: General Relativity incorporates inertia in a purely Newtonian way, without modification. However, it has been known for a long time that an inertial mass is quantized in the form of elementary particles, and that there must be some explanation as to why the individual particles that make up a mass resist any change in velocity. Also, Mach's principle is still not fully incorporated or understood in general relativity. This principle strikes at the very heart of the meaning of motion. For an in depth discussion of Mach's principle and it's relationship to general relativity refer to reference 1.

**EMQG:** Inertia (called quantum inertia) is postulated to be the electrical interaction (of which the details are still unknown) between the electrically charged virtual masseon particles of the quantum vacuum and the accelerated real electrically charged masseon that make up a mass. Although quantum inertia is still a postulate (postulate #3 of EMQG) of EMQG, we are reasonably confident that the details of this process will eventually be found. Mach's principle follows directly from the principles of quantum inertia. The virtual particles of the quantum vacuum **are** Mach's invisible reference frame to gauge accelerated motion (from the point of view of acceleration only). In fact, the state of relative acceleration of a mass particle with respect to the background vacuum particles is the actual **origin** of inertial force, and that this same process is also present in gravitational interactions, and responsible for the magnitude of the gravitational mass! However, the quantum vacuum cannot be used as an absolute reference frame for constant velocity motion, and therefore velocity remains relative in EMQG, just as it is in relativity.

### (D) GENERAL RELATIVITY FAILS TO BE QUANTIZED.

**GENERAL RELATIVITY**: In spite of brilliant attempts at unification of general relativity with quantum theory, there still exists no fully accepted quantum gravity theory. It was known even when Einstein proposed general relativity that matter is quantized, and comes in the form of atoms and molecules. These in turn are quantized in the form of elementary particles. Particles behave according to the rules of quantum field theory. Yet, the energy-momentum-stress tensor of a mass is treated as a classical continuum in general relativity. Forces are also known to originate from particle exchange processes.

Is the gravitational force an exception to this general rule? Furthermore, there is growing evidence that space-time itself is also quantized. Yet, no one has found a consistent theory of quantum gravity that reduces to general relativity in the classical limit, leading one to believe that the foundations of general relativity need to be somehow modified.

**EMQG:** EMQG theory *is* the quantization of general relativity. EMQG is totally based on the quantum particle nature of matter and forces, which behave according to the rules of quantum field theory. The 4D space-time curvature of general relativity is a manifestation of the altered state of the quantum vacuum, both under acceleration and under the influence of a nearby gravitational field. EMQG restores the concept of absolute space and separate absolute time in the form of the inherently quantized CA space and CA time, which is not directly observable with our measuring instruments. This CA space and time is inherently quantized because of the very nature of the cellular automata model, which is the fastest possible parallel computer model. The relativistic 4D Minkowski space-time of general relativity is retained in EMQG as purely the end result of quantum particle interactions. Thus, in EMQG the curved 4D space-time geometry paradigm is replaced by a total, quantum particle paradigm, where the quantum particles live in absolute CA space and CA time.

**(E) WHAT IS THE ORIGIN OF THE STRANGE NATURE OF 4D SPACE-TIME IN GENERAL RELATIVITY**

**GENERAL RELATIVITY**: In general relativity, 4D relative Minkowski space-time appears to behave in a strange way. According to general relativity, the universe is a geometric Riemann 4D curved space-time on <u>all</u> distance scales. Yet, the 4D space-time curvature is not considered absolute for different gravitational observers (or for different accelerated observers) in various states of relative motion. Instead, 4D space-time depends very much on the particular physical circumstance or motion involved. In other words, 4D space-time is relative! Yet the presence of a large mass is supposed to cause an ***absolute amount*** of 4D space-time curvature, depending on the magnitude of the mass. In the case of an accelerated frame, the amount of curvature is also not absolute, but depends only on the amount of acceleration possessed by the observer!

For example, an observer 'A' stationed on the surface of the earth finds himself embedded in 4D, curved space-time. Yet, another observer 'B' in free fall near the earth finds himself in perfectly flat 4D space-time. Let us suppose that observer 'B's free fall path takes him directly past observer 'A', whom he can even momentarily touch as he falls past him. At the moment of contact the space-time is *still different* for *both* observers! This seems to be a very strange way to construct the fundamental geometry of our universe. Even though both observers can actually physically touch each other for a brief moment, each observer still lives in different 4D space-times! What is even stranger about space-time is that observer 'A' has a space-time curvature directed along the radius vector only (as can be seen from the Schwarzchild metric). In other words, the curvature of the space and time components of the metric is directional! No curvature results in directions that are parallel to the surface of the earth (we ignore the earth's surface shape). To add to the confusion, if we imagine that observer 'A' decides to take on a new acceleration (2 g's) parallel to the earth's surface, he now lives in a curved space in both the parallel and perpendicular directions. However, now the curvature is unequal in value in both the parallel and perpendicular directions!

**EMQG:** At the most fundamental distance scales, there exists only absolute CA 3D space, and separate absolute CA time, which are both inherently quantized by the structure of the CA model. This absolute space and time is ***not*** curved in the presence of a gravitational field, nor is space and time unified as it is in relativity! Furthermore, absolute space and time is *not* relative, and the state of motion of an observer does **not** affect this coordinate system at all! The measurable relativistic 4D space-time is one layer above the absolute space consisting of cells, and absolute time consisting of 'clock' cycles. We have seen that the 4D relativistic, curved space-time, which is present in accelerated frames and gravitational frames, is an abstraction. It results from the behavior of light and matter in the presence of the accelerated virtual particles of the quantum vacuum, which can be looked at as a kind of a special 'Fizeau-like' fluid. Generally for observers in gravitational and accelerated frames, the virtual particles are in a state of accelerated motion. Observer 'B' can cancel this curved 4D space-time simply by taking the same acceleration as the virtual particles of the quantum vacuum. This restores the virtual particles of the quantum vacuum (from the point of view of his reference frame) to the equivalent state that the observer would 'see' in flat, special relativistic, 4D space-time. Similarly, observer 'A' senses directional space-time curvature simply because the direction of virtual particle acceleration is directed downward, and not side to side. However, if he decides to accelerate in a direction parallel to the surface of the earth, he will observe the virtual particles of the quantum vacuum accelerating towards him from two different directions, downwards and towards him. In other words, there are two acceleration vectors from the quantum vacuum, one vector directed downwards with an acceleration of 1g, and the other equal in magnitude to the amount of forward acceleration of 2g. This results in unequal space-time curvature in both directions, because the accelerated Fizeau-like quantum vacuum fluid has two different acceleration components in the x and y direction. This introduces two different and independent 4D space-time curvatures.

(F) **GENERAL RELATIVITY PREDICTS SPACE-TIME SINGULARITIES.**

**GENERAL RELATIVITY**: Space-time singularities and black holes are a direct consequence of the mathematics of general relativity theory. However, it is **very** doubtful that these mathematical monsters really exist. In fact, a singularity in physics usually means that a theory has somehow broken down.

**EMQG:** In EMQG, we cannot have a space-time singularity. EMQG is based on the idea of 3D absolute CA space and absolute CA time, that are inherently quantized in the form of cells on a cellular automata. This model prohibits an infinite density of particles to accumulate in one place due to gravitational collapse (and thus predicts that Quantum

Field Theory will break down at very high particle densities). The density of particles that can be supported cannot exceed the number of cells per cubic meter. However, it is not clear what will happen when particle densities become this great.

### (G) GENERAL RELATIVITY FAILS TO ACCOUNT FOR THE VALUE OF THE COSMOLOGICAL CONSTANT

**GENERAL RELATIVITY**: Einstein originally introduced the cosmological constant in 1918, which was a term that he added to his gravitational field equations to make the universe remain in the steady state (or non-expanding, as was believed in his time). This new term corresponded to a nonzero energy momentum tensor of the vacuum. Einstein later abandoned this constant when Hubble discovered in the 1920's that the universe was actually expanding. In fact, Einstein considered the addition of this term to his field equations as the biggest blunder of his life. The cosmological constant represents the measure of the mass-energy density contained in empty space alone. It has been said that the cosmological constant is the *most* striking problem in contemporary fundamental physics. In fact, some theoretical predictions of quantum field theory differs by observations by at least $10^{45}$ and possibly, by 120 orders of magnitude! Experimentally, it is safe to say that its actual value is very close to zero. In fact, S. Hawking once stated that:

*"The cosmological constant is probably the quantity in physics that is most accurately measured to be zero: observations of departures from the Hubble law for distant galaxies place an upper limit of the order of $10^{-120}$."*

On the other hand, quantum field theory predicts that there ought to be plenty of contributions from the virtual particles of the quantum vacuum.

**EMQG:** It is true that the quantum vacuum at any instant of time contains an enormous number of virtual particles, which have both mass and electrical charge. Why is it that their presence is not felt electrically or gravitationally? The cosmological constant is essentially zero because of the symmetrical production of virtual masseon and anti- masseon particle pairs in the quantum vacuum. According to the general principles of quantum field theory, these virtual masseon particle pairs are always created in opposite electrically charged particle pairs, which explains the electrical neutrality of the vacuum. The particle pair creation principle in the quantum vacuum is also applicable to 'gravitational mass charge' as well (postulate #2 of EMQG). Virtual masseon particle pairs are also created in *opposite* gravitational 'mass-charge' pairs. At any instant of time, the quantum vacuum has almost exactly equal numbers of positive and negative electrically charged and 'gravitationally charged' masseon particles, which leads to ***neutrality of both electrical and gravitational forces in the quantum vacuum and a cosmological constant of zero.***

(H) **GENERAL RELATIVITY DOES NOT CONSERVE ENERGY-MOMENTUM.**

**GENERAL RELATIVITY**: It is known that general relativity does not assign an energy-momentum-stress tensor to the gravitational field. There are also serious problems with local energy-momentum conservation. The absence of the energy-momentum-stress tensor is due to the geometric aspect of gravity in general relativity. We know that gravitational waves transfer energy by the rippling of 4D space-time to far away destinations. For example, a pair of orbiting neutron stars emits gravitational waves, which cause them to slowly spiral inwards. This orbital energy loss is precisely the energy carried away by the gravity waves. Thus, gravity waves can do work on a distant mass. However, how can rippling 4D space-time carry energy?

**EMQG:** In general relativity, curved space-time actually does represent the geometry of the universe. This **is** the source of the difficulty in energy conservation in general relativity. In EMQG, curved space-time represents the distorted distribution of the net statistical average acceleration vectors of the virtual electrically charged masseon particles of the quantum vacuum from place to place. These charged vacuum masseon particles interact through electrical forces by photon exchanges with a nearby test mass (which consists of electrically charged masseons). It is this electrical interaction between the quantum vacuum and a test mass that represents the energy interactions in gravitation, and the energy content of curved 4D space-time. The quantum vacuum has a very large energy density, but it is not infinite. Energy and momentum conservation is always obeyed on the quantum particle level in EMQG.

## 3. IMPORTANT APPLICATIONS OF EMQG THEORY

EMQG has been applied to various problems in physics. Here we review some of the more important applications in which EMQG provides the hidden quantum processes behind the phenomena.

**(A) The Quantum Origin of Newton's Laws of Motion**

We are now in a position to understand the quantum nature of Newton's classical laws of motion. According to the standard textbooks of physics Newton's three laws of laws of motion are:

(1) An object at rest will remain at rest and an object in motion will continue in motion with a constant velocity unless it experiences a net external force.
(2) The acceleration of an object is directly proportional to the resultant force acting on it and inversely proportional to its mass. Mathematically: $\Sigma F = ma$, where F and a are vectors.
(3) If two bodies interact, the force exerted on body 1 by body 2 is equal to and opposite the force exerted on body 2 by body 1. Mathematically: $F_{12} = -F_{21}$.

Newton's first law explains what happens to a mass when the resultant of all external forces on it is zero. Newton's second law explains what happens to a mass when there is a nonzero resultant force acting on it. Newton's third law tells us that forces always come in pairs. In other words, a single isolated force cannot exist. The force that body 1 exerts on body 2 is called the action force, and the force of body 2 on body 1 is called the reaction force.

In the framework of EMQG theory, Newton's first two laws are the direct consequence of the (electromagnetic) force interaction of the (charged) elementary particles of the mass interacting with the (charged) virtual particles of the quantum vacuum. Newton's third law of motion is the direct consequence of the fact that all forces are the end result of a boson particle exchange process.

**(1) Newton's First Law of Motion:**

In EMQG, the first law is a trivial result, which follows directly from the quantum principle of inertia (postulate #3). First a mass is at relative rest with respect to an observer in deep space. If no external forces act on the mass, the (charged) elementary particles that make up the mass maintain a *net acceleration* of zero with respect to the (charged) virtual particles of the quantum vacuum through the electromagnetic force exchange process. This means that no change in velocity is possible (zero acceleration) and the mass remains at rest. Secondly, a mass has some given constant velocity with respect to an observer in deep space. If no external forces act on the mass, the (charged) elementary particles that make up the mass also maintain a *net acceleration* of zero with respect to the (charged) virtual particles of the quantum vacuum through the electromagnetic force exchange process. Again, no change in velocity is possible (zero acceleration) and the mass remains at the same constant velocity.

**(2) Newton's Second Law of Motion:**

In EMQG, the second law is the quantum theory of inertia discussed above. Basically the state of *relative* acceleration of the charged virtual particles of the quantum vacuum with respect to the charged particles of the mass is what is responsible for the inertial force. By this we mean that it is the tiny (electromagnetic) force contributed by each mass particle undergoing an acceleration 'A', with respect to the net statistical average of the virtual particles of the quantum vacuum, that results in the property of inertia possessed by all masses. The sum of all these tiny (electromagnetic) forces contributed from each charged particle of the mass (from the vacuum) is the source of the total inertial resistance force opposing accelerated motion in Newton's F=MA. Therefore, inertial mass 'M' of a mass simply represents the total resistance to acceleration of all the mass particles.

**(3) Newton's Third Law of Motion:**

According to the boson force particle exchange paradigm (originated from QED) all forces (including gravity, as we shall see) result from particle exchanges. Therefore, the force that body 1 exerts on body 2 (called the action force), is the result of the emission of force exchange particles from (the charged particles that make up) body 1, which are readily absorbed by (the charged particles that make up) body 2, resulting in a force acting on body 2. Similarly, the force of body 2 on body 1 (called the reaction force), is the result of the absorption of force exchange particles that are originating from (the charged particles that make up) body 2, and received by (the charged particles that make up) body 1, resulting in a force acting on body 1. An important property of charge is the ability to readily emit <u>and</u> absorb boson force exchange particles. Therefore, body 1 is both an emitter and also an absorber of the force exchange particles. Similarly, body 2 is also both an emitter and an absorber of the force exchange particles. This is the reason that there is both an action and reaction force. For example, the contact forces (the mechanical forces that Newton was thinking of when he formulated this law) that results from a person pushing on a mass (and the reaction force from the mass pushing on the person) is really the exchange of photon particles from the charged electrons bound to the atoms of the person's hand and the charged electrons bound to the atoms of the mass on the quantum level. Therefore, on the quantum level there is really is no contact here. The hand gets very close to the mass, but does not actually touch. The electrons exchange photons among each other. The force exchange process works both directions in equal numbers, because all the electrons in the hand and in the mass are electrically charged and therefore the exchange process gives forces that are equal and opposite in both directions.

**(B) The Quantum Physics Behind Gravitational Waves**

The physics of gravitational waves (GW) in EMQG is definitely not the same as the physics of electromagnetic waves described by QED. For a periodically accelerating large mass (or masses), the fluctuating graviton flux is responsible for the periodic disturbance in the net acceleration of the quantum vacuum particles in the immediate vicinity, which will affect another test mass. For example, a close pair of relativistic neutron stars in orbit would both periodically disturb the state of the virtual particles of the quantum vacuum through periodically fluctuating graviton exchanges. Thus, the outward propagating GW is really a time varying periodic increase and decreases in the net acceleration of the virtual particle acceleration vectors with respect to the neutron stars. Once the GW is started, however, it is self sustaining throughout space (as is electromagnetic radiation). The reason that it is self-sustaining is due to the local electromagnetic vacuum process. The strong electromagnetic component in the GW is the gravitational waves primary means of acting upon other masses. The GW periodic excitation propagates electromagnetically, because the quantum virtual particles are constantly redistributing their acceleration vector imbalance among each other via photon exchanges, as the GW travels outward. It is known in the context of general relativity theory that the GW's have a very large stiffness (in analogy with Hooke's law), and thus a large energy density. This fact follows from Einstein's Gravitational Field equations. This is easily explained in EMQG as arising from the very large energy density of the vacuum disturbance itself. The virtual particles of the quantum vacuum have a large particle density that is roughly on the order of $10^{90}$ particles per cubic meter (Plank length cubed presents a rough upper limit to the particle density per cubic meter on a CA, but the exact value is not known). Because of this, the fluctuating quantum vacuum disturbance (GW) carries a large energy density, and is quite capable of explaining the stiffness of the GW wave. In fact, the GW is capable of vibrating a large aluminum cylinder after traveling hundreds of light years in space.

**(C) The Physical Interpretation of the Lense-Thirring Effect on the Quantum Level**

We now apply the principles of EMQG to calculate the amount of inertial frame dragging (Lense-Thirring effect) on the earth. The Lense-Thirring effect is a tiny perturbation of the orbit of a particle caused by the spin of the attracting body, first calculated by the physicists J. Lense and H. Thirring in 1918 using general relativity. Einstein's general relativity predicts the perturbation in the vicinity of the spinning body, but the effect has not been accurately verified experimentally. However, recent work using the LAGEOS and LAGEOS II earth orbiting satellites has rendered an unconfirmed experimental value that agrees with theory to an accuracy of about 20 %. The Lense-Thirring effect has also been interpreted as being due to gravitomagnetic fields (section 19.2 of ref. 1), and also tied in with Mach's principle. It is hoped that with the launch of the Gravity Probe B (co-developed by Stanford University) by NASA the Lense-Thirring effect will be measured to an unprecedented accuracy of 1% or better. The Gravity Probe B (there was a different Gravity Probe A launched earlier by NASA) is a drag free satellite carry an four ultra-precise gyroscopes that will be put in a polar orbit around the earth at a height of about 400 miles.

An important consequence of the Lense-Thirring effect is that the orbital period of a test mass around the earth depends on the direction of the orbit! A test mass that has an orbit which revolves around the earth in the same direction of the spin rotation would be longer then the orbital period of the same test mass revolving opposite to the direction of the spin of the earth. The difference in the orbital period of the two test masses becomes smaller with increasing height until it disappears when the orbits are at infinity. The Lense-Thirring effect can also be be thought of as a kind of 'a dragging of inertial frames' first named by Einstein himself. The Lense-Thirring effect for a rotating mass is most pronounced as the angular velocity of the rotating mass increases. The basic reason for inertial frame dragging is the finite speed of propagation of the graviton particle (the speed of light). This allows time for a large rapidly spinning mass to rotate a small amount while the graviton is still in flight as it propagates outwards. The finite velocity of the graviton particle along with the downward $GM/R^2$ acceleration component of the charged virtual particles of the quantum vacuum is entirely responsible for inertial frame dragging.

We now examine inertial frame dragging for a weak gravitational field such as the earth (the strong field inertial frame dragging is a very formidable problem in EMQG). Gravitons are emitted in huge numbers from the earth. Since we know that gravitons are physically very similar to photons, we can predict the characteristics of this graviton flux. The graviton flux can be visualized as being the same as a rotating flashlight emitting light outwards as it rotates. Since the velocity of the source does not effect the velocity of light, we conclude that the velocity of the spinning earth does not affect the motion of the gravitons. (However, gravitons are affected by the state of accelerated motion of the virtual particles, because gravitons do scatter with the virtual masseon particles of the quantum vacuum. This scattering is small for a mass the size of the earth, but for an object with a very strong gravitational field that is rotating at relativistic speeds, the calculation of the Lense-Thirring effect is an extremely difficult problem in EMQG).

As the gravitons propagate outward, they encounter virtual masseon particles in the quantum vacuum. Since masseons posses 'mass-charge', the vacuum is accelerated down. The virtual masseons are therefore accelerated in the same direction of the motion of the graviton particle flux. The magnitude of the acceleration depends on the graviton flux at a point 'r' from the center, and is given by $GM/r^2$. If a test mass (composed of real electrically charged masseon particles) is dropped onto the surface of the earth, the electromagnetic interactions between the real masseons of the mass and the virtual masseons causes the test mass to accelerate downwards at 1g in the direction of the graviton flux. Thus, from the perspective of an external observer, the mass falls along the radius vectors. The observer on the surface of the equator 'sees' the graviton flux leaving the equator in curved paths he is carried along with the earth's rotation. From his frame of reference the gravitons appear to curve, and the average acceleration vector of the virtual particles follows this same curved path, with the acceleration vectors increasing in magnitude the closer to the center. These curved paths also represent the path of light that light will take if it propagates straight up (in other words, geodesic paths). In absolute CA units, the light velocity varies upwards (or downwards) along these curved paths due to the Fizeau-like scattering with the quantum vacuum, and thus these paths represent the direction of the 4D space-time curvature. These paths are deflected when compared to the non-rotating earth.

The equation of the outward propagating graviton curve (in the equatorial plain) for a clockwise rotating earth turns out to be Archimedes' Spiral, and takes the form $r = k\theta$ in polar coordinates, where k is a constant (r is the distance that the graviton travels). The constant k depends only on the velocity of the graviton 'c' and the velocity $v = 2\pi R/T_p$ of the earth's rotation, where $T_p$ is the rotation period of the earth and R is the earth's radius, and the time t of transit. If k is small then the spiral has a high curvature, and if k is large the curvature is small. The ratio 'c/v' determines the value of the spiral constant k. If c were to be very large, then $k \gg 1$ which causes the gravitons to 'unwind' slowly; and if v were to be very large, then $k \ll 1$ which causes the gravitons to 'unwind' rapidly.

Based on these considerations, the equation for the spiral is:

$r = c^2 \theta/2\pi v$  where $v = 2\pi R/T_p$, and $\theta = 2\pi t v/c$

Let us calculate the shape of the spiral for the observer A on the earth. The earth has a radius $R = 6.37 \times 10^6$ meters, and a rotation period $T_p = 24$ hours. Therefore, the v = 463 m/sec. The equation of the spiral for the earth is: $r = 103,176\ c\theta$. We wish to solve for the angle $\theta$ at the earth's surface, where $r = 6.37 \times 10^6$ meters. Therefore: $\theta = 2.05 \times 10^{-7}$ radians = 42.5 milli-arcseconds. This agrees well with the standard prediction based on general relativity. However, the result is much easier to derive and visualize with EMQG than with general relativity theory. This angle represents the deflection of the downward accelerating virtual particles of the quantum vacuum with respect to the non-rotating earth. Furthermore, the deflection angle varies with height (along a spiral path) which causes inertial frame dragging.

## 4. EXPERIMENTAL TESTS OF EMQG THEORY

EMQG reveals several new experimental tests that give results that are different from conventional general relativistic physics. Although most of these experiments are very difficult to perform, they do provide a solid foundation for testing the principles of EMQG theory.

(i) If two sufficiently large pieces of anti-matter can be manufactured in the distant future to allow measurement of the mutual gravitational interaction, then the gravitational force will be found to be repulsive! The force will be equal in magnitude to $-GM^2/r^2$ where M is the mass of each of the equal anti-matter masses, r is their mutual separation, and G is Newton's gravitational constant). This is in clear violation of the principle of equivalence, since in this case $M_i = -M_g$, instead of masses $M_i = M_g$. Antimatter that is accelerated in far space has the same inertial mass '$M_i$' as ordinary matter, but when interacting gravitationally with another antimatter mass it is repelled ($M_g$). (**Note:** The earth will *attract* bulk *anti-matter* because of the large abundance of gravitons originating from the earth of the type that induce attraction). This means that no violation of equivalence is expected for anti-matter dropped on the earth, where anti-matter falls normally (recall that virtual masseons and anti-masseons are both attracted to the earth (postulate #2). However, an antimatter earth will repel an antimatter mass dropped on the earth. Recent attempts at measuring earth's gravitational force on anti-matter (e.g. anti-protons will not reveal any deviation from the principle of equivalence).

(ii) If an extremely large test mass and a very small test mass are dropped simultaneously on the earth (in a vacuum free of air resistance), there will be an extremely small difference in the arrival time of the masses, in slight violation of the principle of equivalence. This effect is on the order of $\approx \Delta N \times \delta$, where $\Delta N$ is the difference in the number of masseon particles in the two masses, and $\delta$ is the ratio of the gravitational to electric forces for one masseon. This experiment is very difficult to perform on the earth, because $\delta$ is extremely small ($\approx 10^{-40}$), and $\Delta N$ cannot be made sufficiently large. To achieve a difference of $\Delta N = 10^{30}$ particles between the small and large mass requires dropping a molecular-sized atomic cluster and a large military tank simultaneously in the vacuum in order to give a measurable deviation. Note: For ordinary objects that might seem to have a large enough difference in mass (like dropping a feather and a tank), the difference in arrival time may be obscured by background interference, or by quantum effects like the Heisenberg uncertainty principle which restrict the accuracy of time measurements.

(iii) Currently it is well known that graviton particles cannot be detected directly by any laboratory experiments, due to the extremely weak interaction predicted between individual mass particles and the graviton exchange particles. However if in the future gravitons can be somehow detected by the invention of a graviton detector/counter, then there will be experimental proof for the violation of the strong principle of equivalence. The strong equivalence principle states that all the laws of physics are the same for an observer situated on the surface of the earth as it is for an accelerated observer on a rocket (1 g). The graviton detector will find a tremendous difference in the graviton particle counts in these two cases, because gravitons are vastly more numerous here on the earth then in a mass undergoing acceleration.

(iv) EMQG theory predicts that all test masses (which are electrically neutral, overall) interacts with the electrically charged virtual particles falling near the earth, through a strong electrical force component occurring between the electrically charged particles that constitute the mass and the electrically charged virtual particles of the quantum vacuum. Therefore, a sensitive mass measurement of a test mass near the earth might be disrupted by experimentally manipulating the distribution of the electrically charged virtual particles of the nearby quantum vacuum. If a rapidly fluctuating magnetic field (or rotating magnetic fields) is produced under a mass, it might effect the instantaneous virtual charged particle spectrum and disrupt the tiny electrical forces contributed by each electrically charged masseon of the mass. This may reduce the measured gravitational mass of an object in the vicinity (this would also affect the inertial mass). In a sense, this device would act like a primitive weak "anti-gravity" device. The virtual particles are constantly being "turned-over" in the vacuum at different rates depending on the energy, with the high frequency particles (and therefore, high-energy particles) being replaced the quickest. If a magnetic field is made to fluctuate fast enough so that it does not allow the new virtual particle pairs to replace the old and smooth out the disruption, the spectrum of the vacuum will be altered.

According to conventional physics, the energy density of virtual particles is infinite, which means that all frequencies of virtual particles are present. In EMQG there is a definite upper cut-off to the frequency, and therefore the highest energy according to the Plank's law: $E=h\upsilon$, where $\upsilon$ is the frequency that a virtual particle can have. This frequency cutoff is very roughly on the order of the plank distance scale. We can therefore state that the smallest wavelength that a virtual particle can have is on the order of about $10^{-35}$ meters, e.g. the plank wavelength (or a corresponding maximum Plank frequency of about $10^{43}$ hertz for very high velocity ($\approx c$) virtual particles). Unfortunately for our "anti-gravity" device, it is technologically impossible to disrupt the highest frequencies. According to the uncertainty principle, the relationship between energy and time is: $\Delta E \times \Delta t < h$. This means that the high frequency end of the spectrum consists of virtual particles that "turns-over" the fastest. To give measurable mass change the higher frequencies of the vacuum must be disrupted, which requires magnetic fluctuations on the order of at least $10^{20}$ cycles per seconds. Therefore, only lower frequencies virtual particles of the vacuum can be practically affected, and only small changes in the measured mass can be expected with today's technology. As a result of this, a relationship should exist between the amount of gravitational (or inertial) mass loss and the frequency of electromagnetic fluctuation or disruption. The higher the frequency the greater the mass loss.

Work on the Quantum Hall Effect by Laughlin has suggested that the electron density in a two-dimensional sheet under the influence of a strong magnetic field causes the electrons to move in concert, with very high speed swirling vortices created in the resulting 2D electron gas. In ordinary magnetic fields, electrons are merely 'pushed' around, while a strong magnetic field causes the electrons to swirl in high-speed 'whirlpools'. There is also a possibility that this 'whirlpool' phenomena holds for the virtual particles of the quantum vacuum under the influence of a strongly fluctuating magnetic field. These high-speed whirlpools might disrupt the high frequency end of the spectral distribution of electrically charged virtual particles in small pockets. Therefore, there might be a greater mass loss under these circumstances. Recent experiments on mass reduction with rapidly rotating magnetic fields are inconclusive at this time. Reference 30 gives an excellent and detailed review of the various experiments on reducing the gravitational force with superconducting magnets.

## 5. CONCLUSIONS

We have reviewed a new quantum gravity theory called Electro-Magnetic Quantum Gravity which is manifestly compatible with Cellular Automata theory, and is also based on a new theory of inertia proposed by R. Haisch, A. Rueda, and H. Puthoff. EMQG has introduced several new important paradigms in physics:

(1) The universe consisting of matter, energy, forces, space and time is the result of a vast numeric 3D Geometric Cellular Automata. One set of mathematical laws is programmed in each and every cell. The CA model automatically presents the universe with a maximum speed limit for the motion of any information pattern.
(2) Newtonian Inertia is really a hidden quantum vacuum process, which results from the electrical force interactions between the charged virtual fermion particles of the quantum vacuum which are in a state of *relative acceleration* with the electrically charged fermion particles that make up the accelerating mass. The sum of the resulting electrical forces for each particle in the mass is the total inertia of a mass.
(3) The magnitude of the gravitational force of a mass that is stationary on the earth's surface results from exactly the *same* hidden quantum process in (2) above, where now it is the ***quantum vacuum that is accelerating***. The *reversal* of the role of the quantum vacuum and the test mass in inertia and in gravity is ultimately what is responsible for the Einstein Equivalence Principle.
(4) At distance and time scales smaller than the Plank Scale, there exits a kind of a quantized, absolute 3D CA space, and a ***separate***, quantized time (the CA 'clock' cycles). CA space and time are *not* accessible by direct measurement.
(5) Gravitational force interactions result from the activities of ***both*** the photon exchange processes and the graviton exchange processes. Graviton exchanges between the earth and the qunatum vacuum causes the electrically charged virtual fermion particles of the quantum vacuum to fall during their brief lifetimes. Photon exchanges between the electrically charged matter particles and the accelerating quantum vacuum particles causes the bulk of the gravitational force of a test mass (with a negligible gravitational force component that comes from the direct graviton exchanges between the earth and the test mass).

(6) Curved 4D space-time on the earth (and in accelerated rockets) is the consequence of the state of relative accelerated 'flow' of the virtual particles of the quantum vacuum (with respect to matter), which can be thought of as a special 'Fizeau-like fluid' that flows through everything. Like in the Fizeau experiment, this vacuum fluid flow effects the propagation of light, and all space and time measuring instruments.

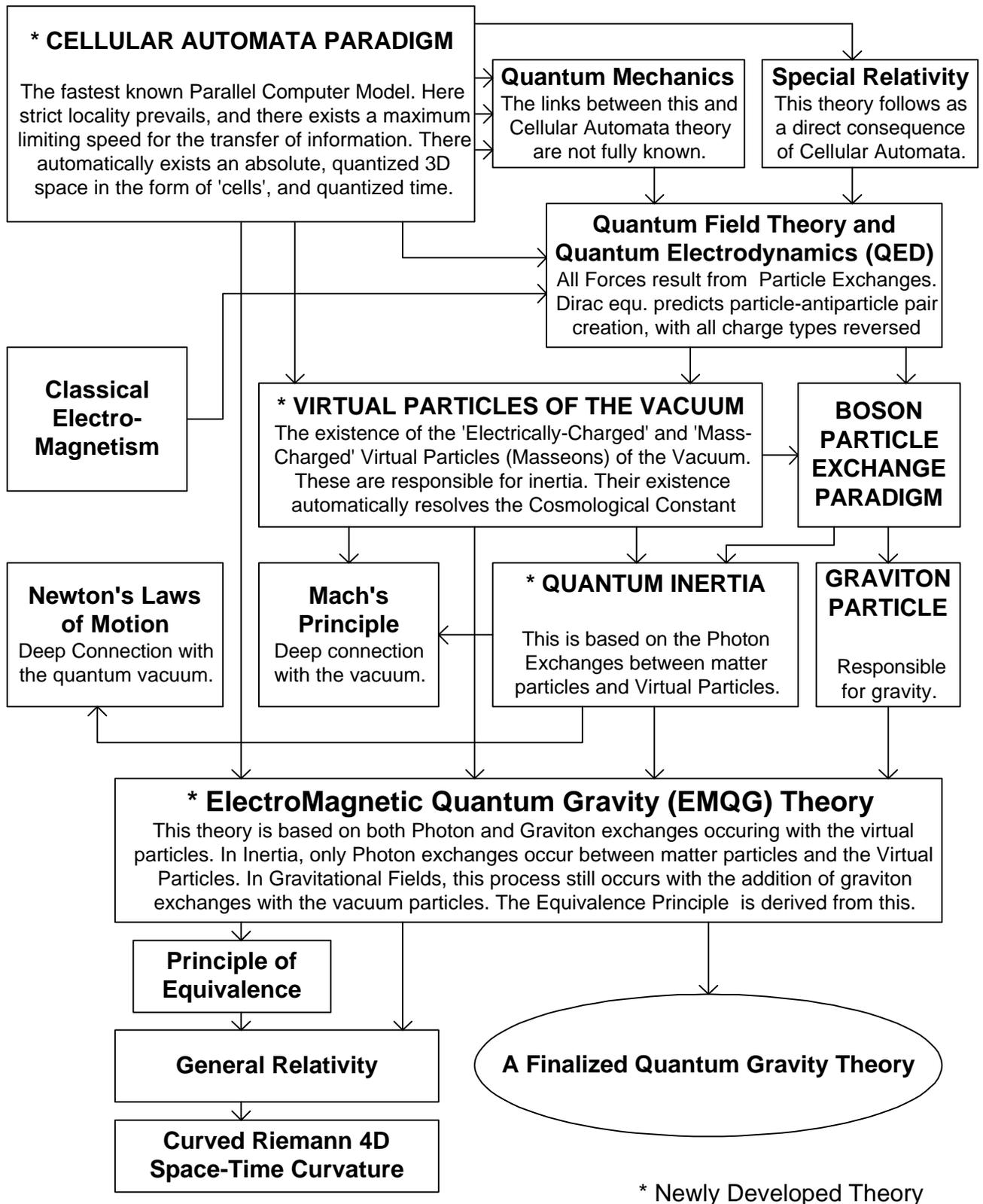

**Figure #1 - BLOCK DIAGRAM OF RELATIONSHIP OF CA AND EMQG WITH PHYSICS**

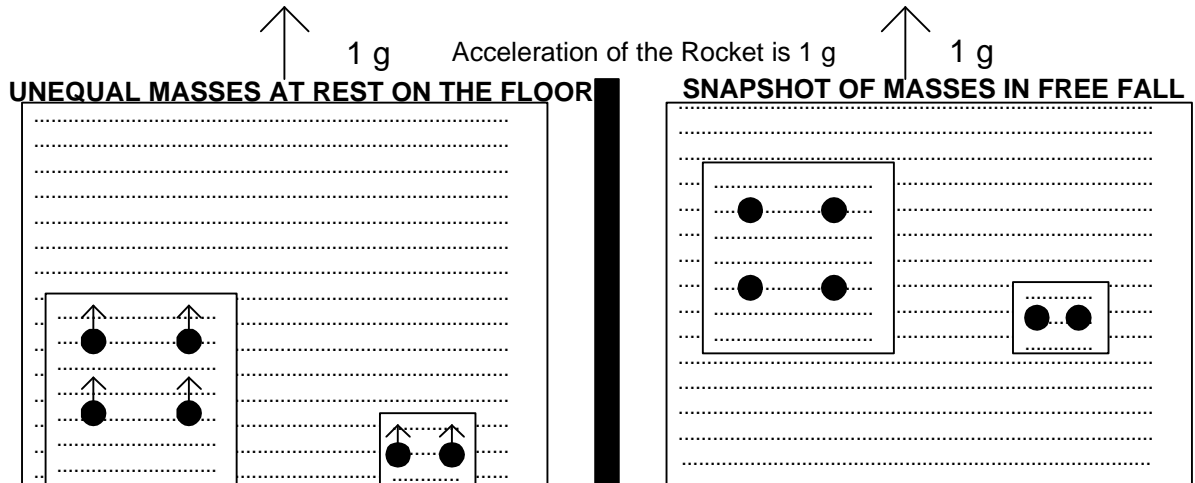

**Figure 2A** - Masses '2M' and 'M' at rest on the floor of the rocket

**Figure 2B** - Masses '2M' and 'M' in free fall inside of a rocket

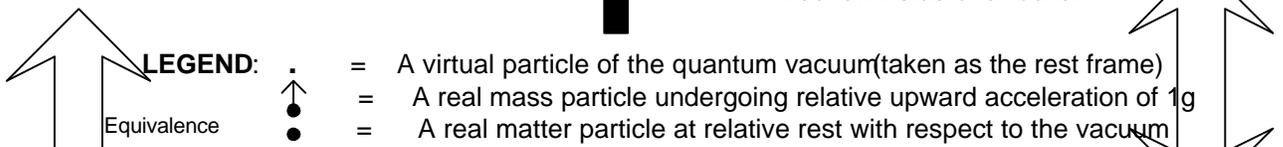

**LEGEND:**
- **.** = A virtual particle of the quantum vacuum (taken as the rest frame)
- (arrow with dot) = A real mass particle undergoing relative upward acceleration of 1g
- **●** = A real matter particle at relative rest with respect to the vacuum

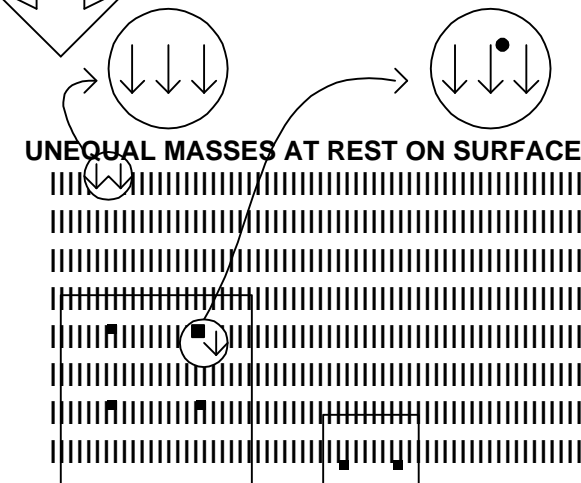

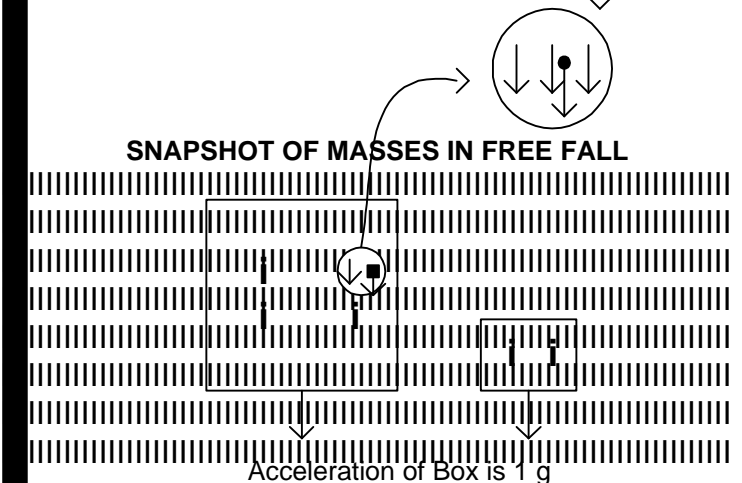

Surface of the Earth where gravity produces a 1 g acceleration

**Figure 2C** - Masses '2M' and 'M' at rest on Earth's surface

**Figure 2D** - Masses '2M' and 'M' in free fall above the Earth

**LEGEND:**
- **l** = Relative downward acceleration (1g) of a virtual particle
- **i** = Relative downward acceleration (1g) of a real matter particle
- **.** = A real stationary matter particle (with respect to the earth's center)

**Figure #2 - SCHEMATIC DIAGRAM OF THE PRINCIPLE OF EQUIVALENCE**